\documentclass[aps,prb,twocolumn,showpacs,superscriptaddress]{revtex4}
\usepackage{amssymb,amsmath}
\usepackage{graphicx}
\usepackage{multirow}
\usepackage{hyperref}
\usepackage{appendix}
\usepackage{mathtools}
\usepackage{amsmath}
\usepackage{xcolor}
\usepackage{siunitx}
\graphicspath{{./IMG/}}

\begin{document}

\title{
Beyond the one-dimensional configuration coordinate model of photoluminescence}

\author{Yongchao Jia}
\email[]{yongchao.jia@uclouvain.be}
\affiliation{European Theoretical Spectroscopy Facility, Institute of Condensed Matter and Nanosciences, Universit\'{e} catholique de Louvain, Chemin des \'{e}toiles 8, bte L07.03.01, B-1348 Louvain-la-Neuve, Belgium}
\author{Samuel Ponc\'{e}}
\affiliation{Department of Materials, University of Oxford, Parks Road, Oxford, OX1 3PH, UK}
\author{Anna Miglio}
\affiliation{European Theoretical Spectroscopy Facility, Institute of Condensed Matter and Nanosciences, Universit\'{e} catholique de Louvain, Chemin des \'{e}toiles 8, bte L07.03.01, B-1348 Louvain-la-Neuve, Belgium}
\author{Masayoshi Mikami}
\affiliation{Functional Materials Design Laboratory, Yokohama R$\&$D Center, 1000,
Kamoshida-cho Aoba-ku, Yokohama, 227-8502, Japan}
\author{Xavier Gonze}
\affiliation{European Theoretical Spectroscopy Facility, Institute of Condensed Matter and Nanosciences, Universit\'{e} catholique de Louvain, Chemin des \'{e}toiles 8, bte L07.03.01, B-1348 Louvain-la-Neuve, Belgium}
\affiliation{Skolkovo Institute of
Science and Technology, Skolkovo Innovation Center, Nobel St. 3, Moscow, 143026, Russia.}

\date{\today}

\begin{abstract}
The one-dimensional configuration coordinate model (1D-CCM) is widely used for the analysis of photoluminescence in molecules and doped solids, and relies on a linear combination of the equilibrium nuclear configurations of ground and excited states. It delivers an estimation of the energy barrier at which ground and excited state curves cross, semi-classically linked to non-radiative transition rate and thermal quenching. To assess its predictive power for the latter properties, we propose a new \textit{optimized configuration path (OCP) method in which} the ground-state and excited-state forces are mixed instead of their configurations. We also define another one-parameter model thanks a double energy parabola hypothesis (DEPH). 
We compare the OCP method and the DEPH reference with the 1D-CCM for three paradigmatic 4f-5d phosphors Y$_3$Al$_5$O$_{12}$:Ce, Lu$_2$SiO$_5$:Ce, and YAlO$_3$:Ce.
We find that the OCP and DEPH methods yield similar results with geometries that have significantly lower ground-state energies than the 1D-CCM for the same 4f-5d energy difference.
However, the OCP method suffers from the appearance of multiple local minima, rendering the clear determination of the optimal geometry very difficult in practice.
Still the OCP method allows one to quantify the deviations from the 1D-CCM, therefore increasing confidence in the lower bound obtained from the DEPH for the 4f-5d crossing barrier, and its comparison with the energy of the auto-ionization thermal quenching mechanism.
We expect the OCP approach to be applicable
to other luminescent materials or molecules. 
\end{abstract}


\maketitle

\section{Introduction}
\label{sec:intro}

Pioneering works to understand the electron-lattice coupling in luminescent materials date back to Condon's study of diatomic molecules, nearly one hundred years ago, in which he developed a one-dimensional configurational coordinate model (1D-CCM),\cite{condon} as an extension of Frank's work on the dissociation of molecules by light absorption.\cite{franck1926} 
Seitz and Mott used the proposed 1D-CCM to study light efficiency and position of color centers in solids,\cite{seitz1938,mott1938} and opened up the field of qualitative analysis of luminescence. 
Later,  quantitative studies were conducted by Huang, Rhys and Pekar in the 1950's. 
They reduced the multi-dimensional configurational coordinate problem to an effective 1D-CCM by assuming that all relevant lattice nuclear coordinates shared a single frequency at which the electronic system exchanged energy by a weak coupling.\cite{huang1950,pekar1954} 
In that case, a single collective displacement connecting the equilibrium ground-state geometry and the equilibrium excited-state geometry is appropriate to represent the luminescence phenomenon. Such 1D-CCM is used in many different contexts to describe radiative and non-radiative recombination.\cite{struck-book,Henderson-book,dorenbos2005}

\begin{figure}
\includegraphics[scale=0.35]{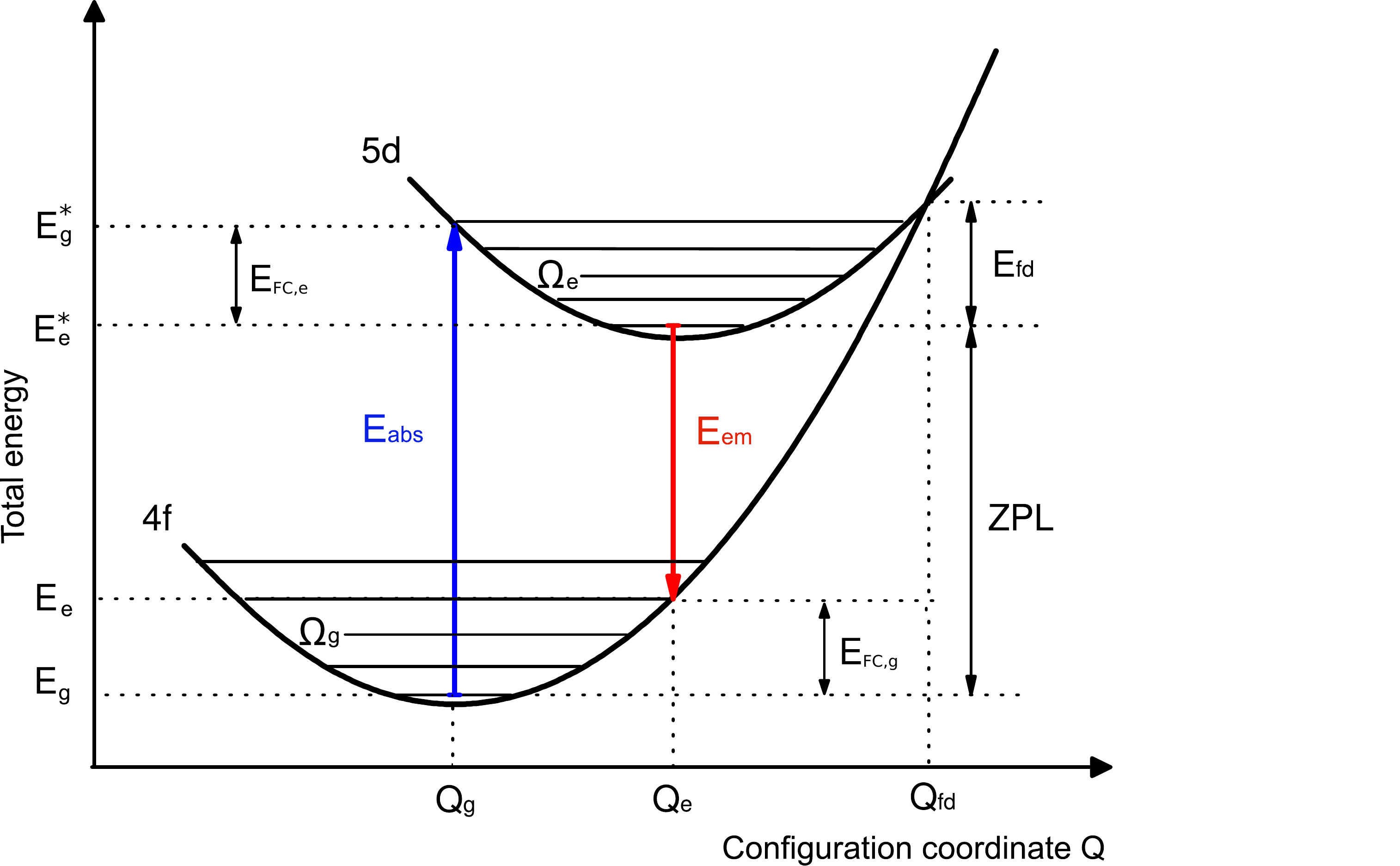}
\caption{The one-dimensional configurational coordinate model in Ce$^{3+}$- or Eu$^{2+}$-doped phosphors, that share 4f-5d excitation. The generalized configuration coordinate $Q$ results from the linear combination of ground state atomic positions (Q$_g$) and excited state ones (Q$_e$). 
Total energies as a function of Q are reported for the electronic ground state (labeled 4f), and for the electronic excited state (labeled 5d), the absorption and emission occur between these energies at E$_g$ and E$_e$, respectively.
The energy lost to the lattice after light absorption defines the Frank-Condon shift E$_{FC,e}$, and similarly E$_{FC,g}$ is the energy lost to the lattice after light emission. Their sum gives the Stokes shift. 
One can associate vibrational frequencies to the ground and excited state curves, $\Omega_g$ and $\Omega_e$. 
They are linked to each spectrum shape through the so-called Huang-Rhys factor $S$, the ratio between a Frank-Condon shift and the related vibrational frequency. ZPL denotes the zero-phonon line, the direct transition energy when no phonon is involved. E$_{fd}$ is the classical activation barrier at atomic positions Q$_{\rm fd}$ for non-radiative recombination through 4f-5d crossing.}
\label{ccd}
\end{figure}

A typical 1D-CCM for rare-earth doped phosphors, like those used in commercial white-LED devices, is shown in Fig.~\ref{ccd}. 
The luminescent lineshape can be obtained by computing the Frank-Condon overlap within the 1D-CCM, a methodology that has been used successfully for decades, even in the context of first-principles approach to the luminescence of solids.\cite{manneback,lax1952,o1953absorption,RevModPhys.31.956,fonger-prb,alkauskas2012,alkauskas2016} 
Such 1D-CCM performs relatively well for the determination of the luminescent lineshape in broad emission materials and gives close agreement with experimental results.\cite{alkauskas2012,alkauskas2016}

In contrast to the study of luminescent lineshape, the performance of the 1D-CCM in predicting the non-radiative recombination has not been the focus of much work,
although the computation of the non-radiative rate has been performed within the 1D-CCM. \cite{Walle2014,Wang2015,Walle2018} 
The 1D-CCM delivers a prediction of the activation energy barrier at the crossing between the ground-state and the excited-state curves, providing a measure of the temperature-dependence of the non-radiative decay rate, a potentially important contribution to thermal quenching in luminescent materials. 
Actually, we could not find any theoretical validation of the 1D-CCM in the context of luminescence efficiency, i.e. the ratio between radiative and non-radiative recombination.  
Most studies rely on a known experimental thermal quenching barrier and there are only a handful works that theoretically predict it in rare-earth doped phosphors, all based on the 1D-CCM.
\cite{dorenbos2005,bachmann2009,qin2017,ponce-2016,Jia-2017,nakano2019}

In this work, we question the predictive power of the 1D-CCM for the activation energy barrier and show indeed that deviations from the
linear combination of equilibrium ground-state and excited-state geometry allows one to decrease the ground-state energy at which some absorption energy is observed. 
We demonstrate this for three paradigmatic Ce$^{3+}$-doped phosphors: Y$_3$Al$_5$O$_{12}$:Ce(YAG:Ce), Lu$_2$SiO$_5$:Ce(LSO:Ce) and YAlO$_3$:Ce(YAP:Ce). 
These materials are widely studied because of their commercial use in white LEDs and as scintillators.\cite{bachmann2009,Jia-2019-2,nikl2000}
The unique 4f$^1$ electron configuration and the fast parity-allowed electric-dipole 4f-5d transition of the Ce$^{3+}$ ions 
allows for the manufacturing of luminescent centers in optical materials with broad-band character.\cite{blasse1967, xie-book,nikl2015,jia-Ce,qin2017} 
To design a commercial product with better performance, full understanding of the physical mechanisms governing luminescence efficiencies is required.

As we did in prior works \cite{Jia-2016,jia-Ce,Jia-2017,Jia-2019-2} and following Fig.~\ref{ccd}, we might assume that both ground-state and excited-state energies are parabolic with respect to some parameter driving a one-dimensional path in the set of configurations. 
This is an ideal situation, that is not met in real materials. 
However, such hypothesis allows one to define a reference behaviour for the energies of the ground and excited states from only four numbers: the energies of the electronic ground and excited states, at the ground-state and excited-state geometries. 
This approach, denoted \textit{double energy parabola hypothesis} (DEPH), makes no assumption on the linearity of the underlying configuration path and therefore goes beyond the 1D-CCM.

To go beyond the parabolic approximation, we propose a new method that we call the \textit{Optimized Configuration Path} (OCP).
This method finds the optimal configuration path such that the atomic positions minimize the ground-state energy, under the constraint of a fixed energy difference between ground 
and excited states. 
This configuration path includes the ground-state and excited-state geometries by construction and also includes the geometry at which the crossing happens. 
This trajectory within the configuration space is non-linear, as the constraint of linear combination of ground and excited state geometries has been relaxed. 
Still, only one continuous parameter connects these geometries.

Nevertheless, we observe a practical difficulty in applying the OCP methodology, due to the appearance of different local minima of the energy functional when the energy difference between the electronic ground-state and the electronic excited state drops below some value
(decrease by about 0.3~eV with respect to the emission energy in the case of the three materials investigated here). 
Indeed, we observe that different geometries that are local minima of the OCP functional can be obtained depending on the starting point for the optimization of this energy functional.
Thus, the unique trajectory corresponding to the global minimum is hard to compute in practice, while including local minima yields bifurcations of trajectories.
We illustrate this problem for the three materials.
Generally speaking, this is a serious concern for the practical usage of the OCP method, 
since global minimization problems have to resort to a different class of algorithms than local minimization problems (e.g. simulated annealing~\cite{Kirkpatrick1983} or Monte Carlo algorithms~\cite{Metropolis1953}, that are much more CPU-time-consuming than conjugate gradient~\cite{Hestenes1952} or Broyden~\cite{Press1988} algorithms, and do not guarantee finding the sought minimum).

By comparing the 1D-CCM, the OCP and the DEPH results, one is nevertheless able to quantify the role of the curvature difference between the ground state and the excited state.
And indeed, such curvature difference differs widely
in our three paradigmatic phosphors .
We will also be able to 
examine the anharmonicities of the different methods by
comparing the OCP and 1D-CCM results with the DEPH ones for the three above-mentioned phosphors.

However, the local minimum problem apparently does not change the assessment of 1D-CCM results
when it comes to associating a lowest ground-state energy with a fixed difference between ground and excited state.
Indeed, in all three materials, the total energies from OCP (even with different local minima) are reasonably consistent with the results of DEPH, which are lower than the result of 1D-CCM. 
On such basis, we confirm our earlier findings that the 4f-5d crossover is not the main thermal quenching mechanism in these phosphors.\cite{ponce-2016,Jia-2017,Jia-2019-2} We believe that such result applies to all Ce$^{3+}$-doped phosphors and possibly to most rare-earth doped phosphors.
 
The paper is structured as follow. 
Sec. \ref{sec:theory} focuses on the theory: we introduce basic definitions, the one-dimensional configuration coordinate model, the optimized configuration path, the double energy parabola hypothesis, and perform a Lagrange parameterization of the double energy parabola hypothesis. 
The first-principles computational methodology and parameters are described in Sec.~\ref{sec:FP}.
In Section \ref{result}, the energy potential landscape for three phosphors is computed using the OCP, the DEPH and the 1D-CCM and compared.
We conclude in Section \ref{conclusion}.  

\section{Theory}
\label{sec:theory}

\subsection{Basic definitions}
\label{sec:basic}

We denote the nuclear positions of the Born-Oppenheimer equilibrium ground state $\mathbf{R}_{\kappa,g}$ and of the equilibrium excited state $\mathbf{R}_{\kappa,e}$ where $\kappa$ labels the different nuclei. 
We suppose that the ground-state Born-Oppenheimer energy can be computed from first principles for any arbitrary nuclear position configuration $\{\mathbf{R}\}$, and is denoted 
$E=E[\{\mathbf{R}\}]$.
Similarly the excited-state Born-Oppenheimer energy is also a function of the nuclear position configuration and is denoted $E^*=E^*[\{\mathbf{R}\}]$. 

Using the definitions introduced in Fig.~\ref{ccd} we have
\begin{align}\label{eq:defsE}
E_{\rm g}   =& E[\{\mathbf{R}_{\rm g}\}] \\
E_{\rm e}   =& E[\{\mathbf{R}_{\rm e}\}] \\
E^*_{\rm g} =& E^*[\{\mathbf{R}_{\rm g}\}] \\
E^*_{\rm e} =& E^*[\{\mathbf{R}_{\rm e}\}],
\end{align}
with the absorption and emission energies given by
\begin{align}\label{eq:absemE}
E_{\rm abs} =& E^*_{\rm g}-E_{\rm g} \\
E_{\rm em}  =& E^*_{\rm e}-E_{\rm e},
\end{align}
and the Frank-Condon shifts given by 
\begin{align}\label{eq:EFC}
E_{\rm FC,e} =& E^*_{\rm g}-E^*_{\rm e} \\
E_{\rm FC,g} =& E_{\rm e}-E_{\rm g}.
\end{align}
Note that out of these four quantities,
only three are independent, as 
\begin{equation}\label{eq:constrainE}
E_{\rm abs}=E_{\rm FC,g}+E_{\rm em}+E_{\rm FC,e}.
\end{equation}
The Born-Oppenheimer equilibrium configurations are mathematically defined as   
\begin{align}\label{eq:Req}
\{\mathbf{R}_{\rm g}\} =& \arg \min E[\{\mathbf{R}\}] \\
\{\mathbf{R}_{\rm e}\} =& \arg \min E^*[\{\mathbf{R}\}],
\end{align}
supposing that $E$ and $E^*$ are convex functions of their argument in the relevant neighborhood of the equilibrium configurations. 
The force on every nucleus, defined as the derivative of the Born-Oppenheimer energy with respect to infinitesimal displacements (for the ground state or the excited state) vanishes, since the equilibrium geometries minimize the corresponding Born-Oppenheimer energy:
\begin{align}\label{eq:Feq}
\mathbf{F}_{\kappa}|_{\{\mathbf{R}_{\rm g}\}} =& -\nabla_{\kappa} E|_{\{\mathbf{R}_{\rm g}\}}=0, \\
\mathbf{F}^*_{\kappa}|_{\{\mathbf{R}_{\rm e}\}}   =& -\nabla_{\kappa} E|_{\{\mathbf{R}_{\rm e}\}}=0.
\end{align}

\subsection{The one-dimensional configuration coordinate model}
\label{sec:1DCCM}

In the 1D-CCM, the equilibrium positions are combined linearly, and are functions of the configuration coordinate $Q$ as follows:
\begin{equation}\label{eq:RQ}
\mathbf{R}_\kappa(Q) = (\mathbf{R}_{\kappa, \rm e}-\mathbf{R}_{\kappa, \rm g}) \frac{Q-Q_{\rm g}}{Q_{\rm e}-Q_{\rm g}}+\mathbf{R}_{\kappa, \rm g}.
\end{equation} 
In the context of the determination of the Huang-Rhys factor and luminescent spectrum shape, the coordinate Q is normalized by including the mass of the atoms, see e.g. Ref.~\onlinecite{alkauskas2012}. 
In the present context, we prefer to normalize it differently, and we introduce the $x$ coordinate, with 
\begin{equation}\label{eq:x}
x=\frac{Q-Q_{\rm g}}{Q_{\rm e}-Q_{\rm g}},
\end{equation}
so that the 1D-CCM nuclei coordinates are  
\begin{equation}\label{eq:Rx}
\mathbf{R}^{\rm 1D}_\kappa(x) = (\mathbf{R}_{\kappa,\rm e}-\mathbf{R}_{\kappa,\rm g})x +\mathbf{R}_{\kappa,\rm g},
\end{equation}
that is, for $x=0$ one gets the ground-state equilibrium nuclear positions
$\mathbf{R}_{\kappa,\rm g}=\mathbf{R}^{\rm 1D}_{\kappa}(x=0)$, while
for $x=1$ one gets the excited-state equilibrium nuclear positions
$\mathbf{R}_{\kappa,\rm e}=\mathbf{R}^{\rm 1D}_{\kappa}(x=1)$.
For sake of simplicity, we have used the superscript 1D instead of the full label 1D-CCM.

The 1D-CCM ground-state $E^{\rm 1D}$ and excited-state $E^{* \rm 1D}$ energy curves as a function of $x$ are computed from
\begin{align}\label{eq:E1Dx}
E^{\rm 1D}(x)  =& E[\{\mathbf{R}^{\rm 1D}(x)\}]  \\
E^{* \rm 1D}(x)=& E^{*}[\{\mathbf{R}^{\rm 1D}(x)\}]. 
\end{align}

For increasing positive $x$, the $E^{1D}(x)$ and $E^{*1D}(x)$ curves might cross. 
We define $x^{\rm c}$ as the first value of $x$ that fulfills 
\begin{equation}\label{eq:DeltaE}
\Delta E=E^{*\rm 1D}(x)-E^{\rm 1D}(x) = 0
\end{equation}
for increasing values of $x$. 
In particular, the energy barrier $E_{\rm b}$ at the crossing $\Delta E=0$ is 
\begin{equation}
E_{\rm b} = E^{\rm 1D}(x^{\rm c})-E^*_{\rm e} =  E^{*\rm 1D}(x^{\rm c})-E^*_{\rm e}.
\end{equation} 

When $\Delta E$ is sufficiently smaller than a typical phonon frequency, there is a non-negligible likelihood of non-radiative recombination through multiphonon emission.
Struck and Fonger~\cite{fonger-1,fonger-2,fonger-3} have explored in considerable details the non-radiative recombination within the 1D-CCM.

However, due to quantum non-adiabatic effects, the relevant energy difference governing non-radiative recombination is slightly lower than the classical exact crossing. 
At some point, the adiabatic and so-called diabatic curves will noticeably differ, indicating that the matrix elements of the electron-nuclei interaction become non-negligible.
There is an ample literature about (non-)crossing effects, see e.g. Refs.~\onlinecite{Fulton1961, McKemmish2011, Bersuker2013} among others.
To avoid such concern, we will analyze the case of ground-state and excited-state energy differing by a given, constrained, energy $\Delta E$, and still consider the adiabatic energies. 

\subsection{Optimized configuration path}
\label{sec:OCP}

The classical transition-state theory~\cite{Glasstone1941} reveals the importance of the size of the energy barrier, governing to a large extent the non-radiative recombination rate through the Arrhenius factor $\exp(\frac{-E_{fd}}{k_BT })$.
We note, however, that the complete picture of non-radiative recombination also includes a prefactor and quantum corrections~\cite{Fermann2000}.
In this work, we choose to focus on finding a lower bound to $E_{fd}$ only. 

Mathematically, the problem of finding the lowest energy $E[\{\mathbf{R}\}]$ at which the difference with $E^{*}[\{\mathbf{R}\}]$ is equal to some fixed value $\Delta E$, in the space of all configurations $\{\mathbf{R}\}$, is expressed as: 
\begin{equation}\label{eq:EOCP}
E^{\rm OCP}(\Delta E)=\min_{\{\{\mathbf{R}\} | E^*[\{\mathbf{R}\}]=E[\{\mathbf{R}\}]+\Delta E\}} E[\{\mathbf{R}\}],
\end{equation} 
 which yields naturally the $\Delta E$-dependent optimized configuration path,
\begin{equation}\label{eq:ROCP}
\{\mathbf{R}^{\rm OCP}(\Delta E)\}= \underset{\{\{\mathbf{R}\} | E^*[\{\mathbf{R}\}]=E[\{\mathbf{R}\}]+\Delta E\}}{\arg \min} E[\{\mathbf{R}\}].
\end{equation}
The same optimized configuration path would be obtained
by similarly optimizing the excited-state energy in the space of constrained configurations. Indeed, their difference being constant, the configuration that minimizes one also minimizes the other.

To find such path in the configuration space, constrained to have a specific energy difference, one can rely on the Lagrange multiplier technique.
The constrained minimization in Eq.~\eqref{eq:EOCP} is indeed equivalent to the unconstrained minimization of 
\begin{multline}\label{eq:EOCPLagrange}
\tilde{E}^{\rm OCP}(\Lambda) =   \\
\min_{\{\mathbf{R}\}} \Big\{ E[\{\mathbf{R}\}] + \Lambda \big(E^*[\{\mathbf{R}\}]-E[\{\mathbf{R}\}]-\Delta E \big) \Big\} 
\end{multline} 
followed by the search for the value of the Lagrange multiplier $\Lambda$ that delivers the sought $\Delta E$, as usual in the Lagrange multiplier approach. 
In what follows, a function of $\Lambda$ will be denoted with a tilde, see the left-hand side of Eq.~(\ref{eq:EOCPLagrange}). 
Such function can be back-transformed to a function of $\Delta E$ in the zone where a one-to-one correspondence exists between these quantities.

For $\Lambda=0$, Eq.~\eqref{eq:EOCPLagrange} reduces to the minimization of the ground-state energy $E[\{\mathbf{R}\}]$, in the space of all configurations, which yields simply the ground-state configuration. 
Thus the latter belongs to the optimized constrained path, with $\Delta E=E_{\rm abs}$. 

For $\Lambda=1$ the minimization of $E^*[\{\mathbf{R}\}]$ is performed (apart the constant $\Delta E$), which delivers the configuration $\{\mathbf{R}_{\rm e}\}$, and yields
$\Delta E=E_{\rm em}$, thus the excited-state configuration belongs to the optimized constrained path as well. 
Still, this path in configuration space does not reduce to the 1D-CCM set of configurations, as it also contains the configuration at which the lowest crossing appears, which is only exceptionally present in the 1D-CCM. 

The configuration that minimizes Eq.~(\ref{eq:EOCPLagrange})  can be determined by computing the forces on the nuclei coordinates, at fixed $\Lambda$. We define 
\begin{equation}\label{eq:F}
\tilde{\mathbf{F}}_{\kappa}^{\rm OCP}(\Lambda)|_{\{{\mathbf{R}}\}} = -(1-\Lambda) \nabla_{\kappa} E|_{\{{\mathbf{R}}\}}
- \Lambda \nabla_{\kappa} E^*|_{\{{\mathbf{R}}\}},
\end{equation}
which yields the condition
\begin{equation}\label{eq:0FOCP}
0= \tilde{\mathbf{F}}_{\kappa}^{\rm OCP}(\Lambda)|_{\{\tilde{\mathbf{R}}^{\rm OCP}(\Lambda)\}}
\end{equation}
that defines the configuration path $\{\tilde{\mathbf{R}}^{\rm OCP}(\Lambda)\}$.  The ground-state and excited-state forces are mixed instead of their configurations like in the 1D-CCM.
Alternatively, this path might be backtransformed as a function of $\Delta E$, giving the solution of Eq.~(\ref{eq:ROCP}).

In Eq.~\eqref{eq:F}, we see that the forces acting on the nuclei configuration, at fixed $\Lambda$, are a simple linear combination of the forces from the ground state and excited state for this nuclei configuration.
The implementation of the search for the OCP is thus rather easy: total energy calculations for the ground state and excited state must simply be coupled, delivering forces as usual, while the optimization of the configuration at fixed $\Lambda$ can be done by combining the computed ground-state and excited-state forces, then using standard optimization algorithms, like Broyden or conjugate gradient~\cite{Abinit2009,Gonze2016}.

\subsection{Double energy parabola hypothesis}
\label{sec:DEPH}
 
We now analyze the consequences of the hypothesis that both the ground-state and excited-state curves are parabolic (DEPH) for a given path of configurations. 
Such hypothesis allows one to compute an approximate energy barrier for crossing, based on the knowledge
of the absorption and emission energies and the Frank-Condon energy shifts only. 

In our earlier work on phosphors~\cite{Jia-2017}, 
such prediction was mentioned to originate from a simplification of the 1D-CCM, and was indeed considered in this context only. Actually, there is no need to rely on the 1D-CCM to examine the consequences of DEPH.
On the contrary, the equivalence of the results obtained for some configuration path
to the results obtained from the DEPH allows one to  characterize the paths for which non-parabolic effects are the smallest.  

Let us consider a set of configurations $\{\mathbf{R}(\lambda)\}$, parameterized by some variable $\lambda$. 
Suppose that the positions of the nuclei are continuous as a function of $\lambda$, with 
$\mathbf{R}_{\kappa,\rm g}=\mathbf{R}_{\kappa}(\lambda=0)$
(the ground-state energy is minimal at this configuration)
and $\mathbf{R}_{\kappa,\rm e}=\mathbf{R}_{\kappa}(\lambda=1)$
(the excited-state energy is minimal at this configuration).
If for such set of configurations, both energy curves can be approximated by parabolas in a sufficiently large range of values of $\lambda$, then one can obtain the crossing energy barrier (provided the corresponding configurations are in this range).

Indeed, the DEPH yields
\begin{align} \label{eq:Elambda}
E(\lambda) &= \lambda^2 E_{\rm FC,g} + E_{\rm g},  \\
\label{eq:E*lambda}
E^*(\lambda) &= (1-\lambda)^2 E_{\rm FC,e} + E^*_{\rm e}.
\end{align} 
The difference between the two energy functions $E^*(\lambda)-E(\lambda)$ is 
\begin{equation}\label{eq:Delambda}
\Delta E = \lambda^2 \Delta C - 2\lambda E_{\rm FC,e} + E_{\rm abs},
\end{equation}
where we have defined the change of curvature between the ground and excited state
\begin{equation}
\label{eq:deltaC}
\Delta C = E_{\rm FC,e} - E_{\rm FC,g}.
\end{equation}
Eq.~\eqref{eq:Delambda} can be inverted (unless $\Delta E$ is bigger than $\Delta C E_{\rm abs}-E_{\rm FC,e}^2$) so that $\lambda$ is obtained as a function of $\Delta E$, in two equivalent formulas:
\begin{align}\label{eq:lambda}
\lambda =& \frac{E_{\rm FC,e}}{\Delta C}
\Bigg[1-\sqrt{1- \frac{\Delta C(E_{\rm abs}-\Delta E)}{E_{\rm FC,e}^2}} \Bigg] \\
 \label{eq:lambda2}
=& 1+ \frac{E_{\rm FC,g}}{\Delta C}
\Bigg[1-\sqrt{1- \frac{\Delta C (E_{\rm em}-\Delta E)}{E_{\rm FC,g}^2}} \Bigg].
\end{align}

The first expression shows that $\lambda=0$ when $\Delta E=E_{\rm abs}$, while the second one yields
$\lambda=1$ for $\Delta E=E_{\rm em}$. 
The excited-state energy is then obtained as a function of $\Delta E$, using Eqs.~\eqref{eq:E*lambda} and \eqref{eq:lambda2}
\begin{multline}\label{eq:E*DeltaE2}
E^*(\Delta E) =  E^*_{\rm e} 
- \frac{E_{\rm FC,e}(E_{\rm em}-\Delta E)}{\Delta C} \\
+ \frac{2  E_{\rm FC,g}^2 E_{\rm FC,e}}{\Delta C^2} 
\Bigg[1-\sqrt{1- 
 \frac{\Delta C (E_{\rm em}-\Delta E)}{E_{\rm FC,g}^2}}\Bigg], 
\end{multline}
which gives the energy barrier
\begin{align}\label{eq:Eb}
E_{\rm b} =& E^*(\Delta E=0)-E^*_e \\ \label{eq:Eb2}
          =& \frac{E_{\rm FC,e}E_{\rm em}}{\Delta C}
\Bigg[ \frac{2  E_{\rm FC,g}^2}{\Delta C E_{\rm em}} 
\Bigg[1-\sqrt{1- 
 \frac{\Delta C E_{\rm em}}{E_{\rm FC,g}^2}}
 \Bigg] -1 \Bigg].
\end{align}

In case $\Delta C=0$, the Franck-Condon shifts are identical and Eq.~\eqref{eq:Delambda} simplify to
\begin{equation}\label{eq:lambda0}
\lambda= 
 \frac{(E_{\rm abs}-\Delta E)}{2E_{\rm FC}}
 =1+ \frac{(E_{\rm em}-\Delta E)}{2E_{\rm FC}},
\end{equation}
leading to 
\begin{equation}\label{eq:E*DeltaE3}
E^*(\Delta E) =  E^*_{\rm e}
+\frac{(E_{\rm em}-\Delta E)^2}{4E_{\rm FC}},
\end{equation}
and the energy barrier
\begin{equation}\label{eq:E*DeltaEb}
E_{\rm b} = \frac{E_{\rm em}^2}{4E_{\rm FC}}.
\end{equation}
Also, in this case,
\begin{equation}\label{eq:EDeltaE0}
E(\Delta E) =  E_{\rm g} 
+\frac{(E_{\rm abs}-\Delta E)^2}{4E_{\rm FC}}.
\end{equation}

In what follows, we will not only focus on the energy barrier Eq.~\eqref{eq:Eb2}, but on functions of $\Delta E$. In this respect, we complete the set of equations by the expression for the ground-state energy as a function of $\Delta E$:
\begin{multline}\label{eq:EDeltaE}
E(\Delta E) =  E_{\rm g} 
- \frac{E_{\rm FC,g}(E_{\rm abs}-\Delta E)}{\Delta C} \\
+ \frac{2  E_{\rm FC,e}^2 E_{\rm FC,g}}{\Delta C^2} 
\Bigg[1-\sqrt{1- 
 \frac{\Delta C(E_{\rm abs}-\Delta E)}{E_{\rm FC,e}^2}}\Bigg]. 
\end{multline}
which can be linked coherently to the excited-state energy Eq.~\eqref{eq:E*DeltaE2}, yielding correctly
\begin{equation}\label{eq:E*DeltaE}
E^*(\Delta E) =  E(\Delta E)+\Delta E.
\end{equation}

In Ref.~\onlinecite{Jia-2017}, an alternative expression for the energy barrier was obtained by
solving Eq.~\eqref{eq:Delambda} for $1/\lambda$, namely
\begin{equation}\label{Ebinverse}
E_{\rm b} = E_{\rm FC,g}\Bigg[1-\frac{E_{\rm abs}}{E_{\rm FC,g} + \sqrt{E_{\rm FC,g}^2 - E_{\rm abs}\Delta C}}\Bigg]^2. 
\end{equation} 
This expression has the advantage not to diverge when $\Delta C=0$, but the corresponding expressions for $E^*(\Delta E)$ and $E(\Delta E)$ are more cumbersome.

\subsection{Lagrange parameterization of the double energy parabola hypothesis}
\label{sec:LDEPH}

In order to compare the results of the OCP with those from the DEPH, we insert Eqs.~\eqref{eq:Elambda} and \eqref{eq:E*lambda} as ground-state and excited-state energies in Eq.~\eqref{eq:EOCPLagrange}, 
\begin{equation}\label{eq:EDEPHLagrange}
\tilde{E}^{\rm DEPH}(\Lambda)=  
\min_{\lambda} \Big\{ E(\lambda) \!+\! \Lambda \big( E^*(\lambda)-E(\lambda)-\Delta E \big)\Big\},
\end{equation} 
and obtain the relationship between $\lambda$ and $\Lambda$,
\begin{equation}\label{eq:relation}
\lambda = \frac{\Lambda}{1-\Delta C (1-\Lambda)/E_{\rm FC,e}}.
\end{equation}
When $\Lambda=0$, this gives $\lambda=0$ and when $\Lambda=1$,
$\lambda=1$ as well.
The function $\Delta \tilde{E}(\Lambda)$ is obtained by combining Eqs.~\eqref{eq:Delambda} and~\eqref{eq:relation},
\begin{equation}\label{eq:DEDEHPLambda}
\Delta \tilde{E}(\Lambda) = E_{\rm abs} - \Lambda \frac{\big( E_{\rm FC,g} (2-\Lambda)+E_{\rm FC,e} \Lambda \big)}
{\big(1-\Delta C(1-\Lambda)/E_{\rm FC,e}\big)^2}.
\end{equation}

This also delivers the ground and excited states from DEPH as a function of $\Lambda$:
\begin{align}\label{eq:EDEHPLambda}
\tilde{E}(\Lambda) =& E_{\rm g} + E_{\rm FC,g} \Bigg(\frac{\Lambda}{1-\Delta C(1-\Lambda)/E_{\rm FC,e}}
\Bigg)^2, \\ \label{eq:E*DEHPLambda}
\tilde{E}^*(\Lambda) =& E_{\rm e}^* + E_{\rm FC,e} \Bigg(\frac{(1-\Lambda)(1-\Delta C/E_{\rm FC,e})}{1-\Delta C(1-\Lambda)/E_{\rm FC,e}}\Bigg)^2.
\end{align}

\begin{figure*}[hbtp]
\includegraphics[width=\textwidth]{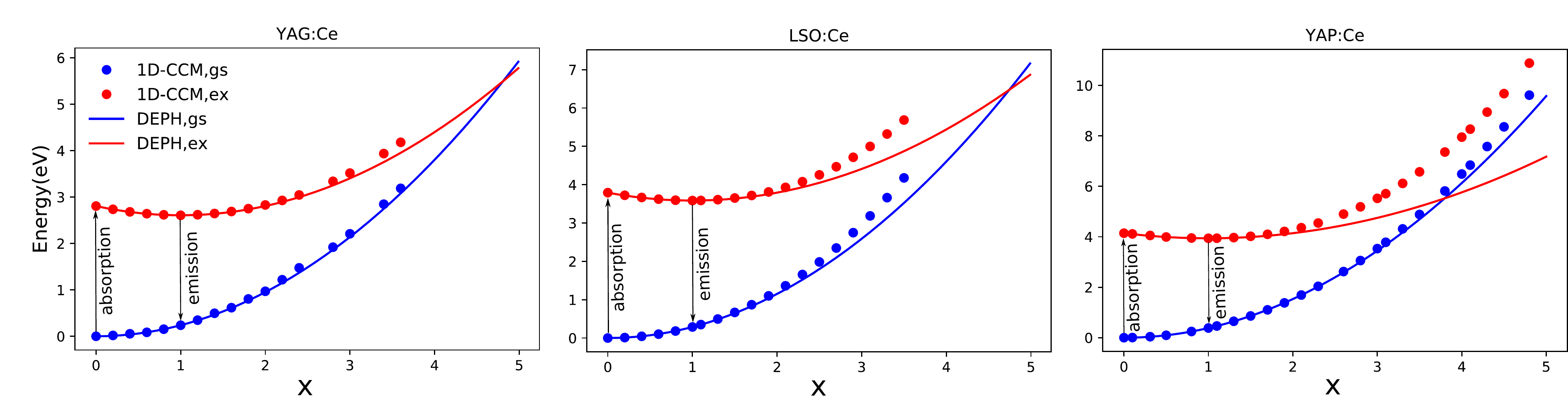}
\caption{The ground-state and excited-state total energies as a function of x, from the 1D-CCM method and from the DEPH, in three Ce$^{3+}$-doped phosphors.(a) YAG:Ce (b) LSO:Ce; (c) YAP:Ce. }\label{figure_1D}
\end{figure*}

  \begin{figure*}[hbtp]
  \centering
 \includegraphics[scale=0.5]{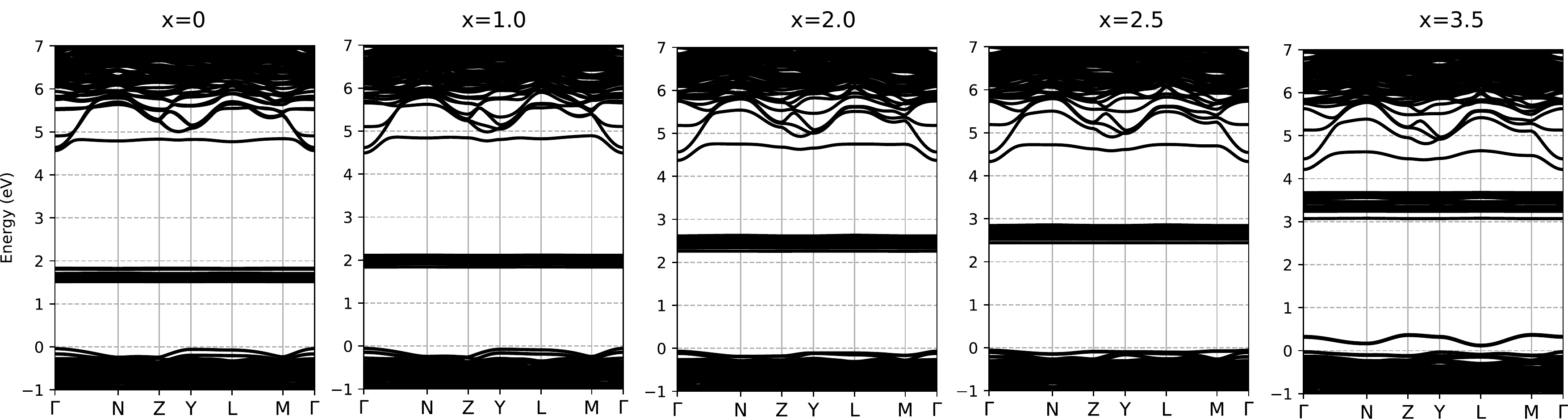}
 \caption{Electronic band structure of LSO:Ce in the excited state, with different $x$ values for the linear combinations of Q$_g$ and Q$_e$ geometry from 1D-CCM.}
 \label{LSO-1D}
 \end{figure*} 
 
\begin{figure*}
\includegraphics[scale=0.44]{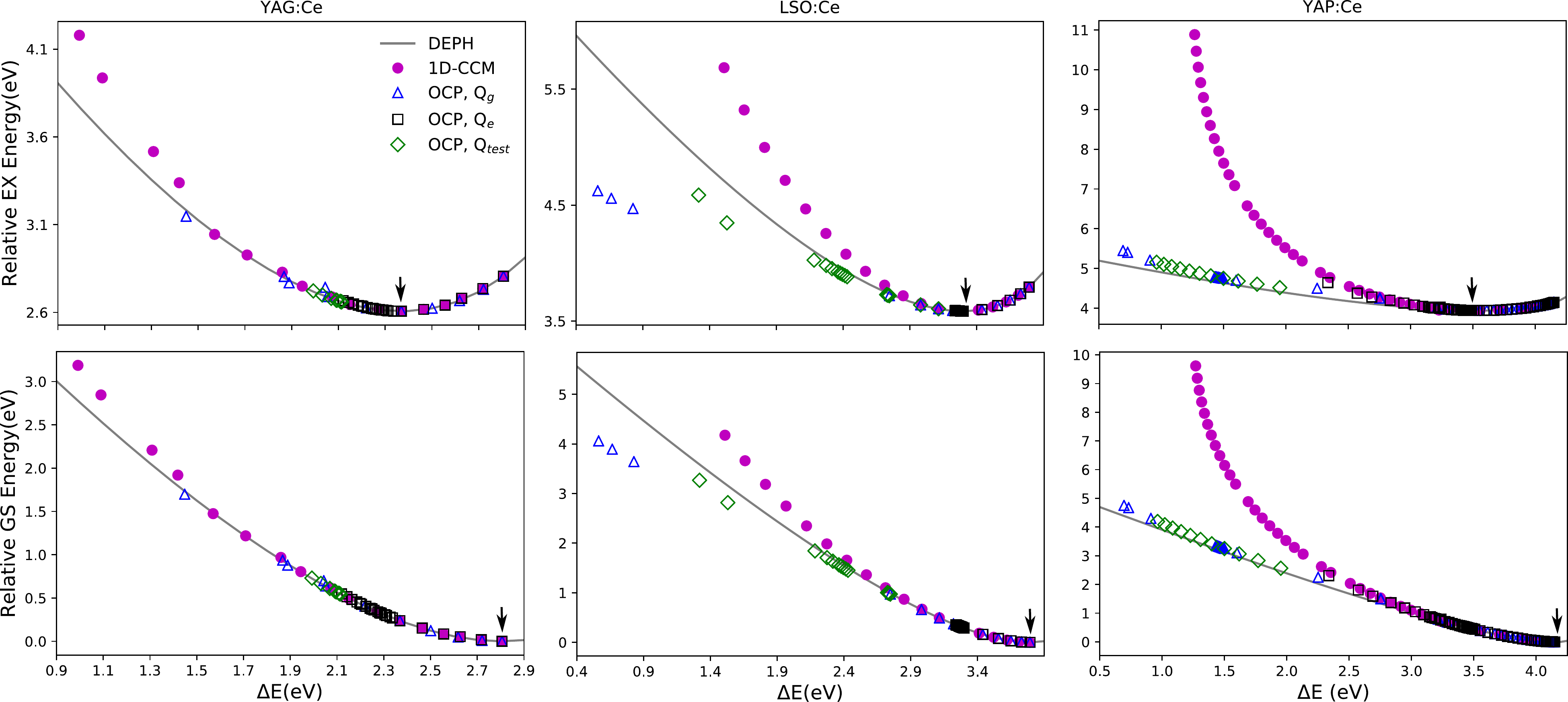}
\caption{The ground-state (lower part) and excited-state (upper part) energy as a function of their difference $\Delta$E, from the 1D-CCM, the OCP method and the DEPH, in three Ce$^{3+}$-doped phosphors: (a) Y$_3$Al$_5$O$_{12}$:Ce (on the left) (b) Lu$_2$SiO$_5$:Ce (center); (c) YAlO$_3$:Ce (on the right). 
The 4f-5d crossing corresponds to $\Delta$E=0, which is never attained, see text, although clear trends appear for decreasing values of $\Delta$E. The reference DEPH continuous black line corresponds to Eq.~(\ref{eq:EDeltaE}) with three parameters (per material) determined by first-principles data. 
The violet full circles are obtained from the one-dimensional configuration coordinate model. 
The other symbols show computed values from the OCP, including possibly different local minima values for $\Delta$E lower than some threshold:
the blue empty triangles are obtained from OCP relaxations starting at the ground-state geometry; the black empty squares are obtained from OCP relaxations starting at the excited-state geometry; the green empty diamond are obtained from test OCP relaxations starting at the OCP geometries with $\Lambda$=1.4, 1.3 and 1.4 for Y$_3$Al$_5$O$_{12}$:Ce, Lu$_2$SiO$_5$:Ce and YAlO$_3$:Ce, respectively. The arrow indicates the minimum energy point for the ground and excited states.
}\label{figure_Delta}
\end{figure*}

\section{First-principles calculations}
\label{sec:FP}

\begin{figure*}
\includegraphics[width=\textwidth]{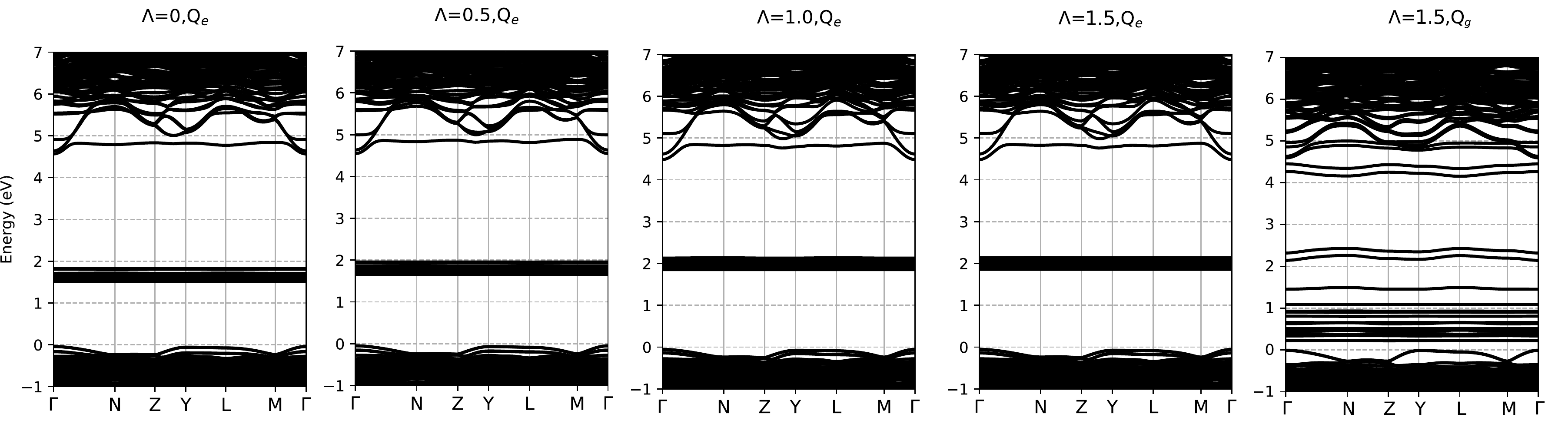}
\caption{Electronic band structure of LSO:Ce in the excited state, for geometries deduced from the OCP method, for four different $\Lambda$ values ($\Lambda=0, 0.5, 1.0, 1.5)$. 
Using the Q$_e$ geometry as input (four leftmost panels), a nearly flat level at 4.8 eV with predominant 5d character can be identified. 
The 4f state eigenenergies (around 2eV) increase slightly with $\Lambda$.
If the OCP optimization at $\Lambda=1.5$ is started from the Q$_g$ geometry, another geometry is found, for which the electronic structure is quite different, see the rightmost panel.
}
\label{figure_bs}
\end{figure*}

In our previous works\cite{jia-Ce,Jia-2019-2,Abinit2009,Gonze2016}
we already examined the three paradigmatic materials YAG:Ce, LSO:Ce and YAP:Ce, among others. 
We performed detailed calculations of absorption energies, emission energies, and Franck-Condon shifts, using the ABINIT software package~\cite{Abinit2009,Gonze2016}. 
In the present work, we stick to the same computational and technical choices.
For sake of completeness, let us mention the key calculation parameters involved. The calculations were performed within density functional theory (DFT) using the projector augmented wave(PAW) method. Exchange-correlation (XC) effects were treated within the generalized gradient approximation (GGA). All of the PAW atomic data sets were directly taken
from the ABINIT website. With these PAW atomic data sets, we performed the structural relaxation and band structure calculations. The
convergence criteria have been set to 10$^{-5}$ Ha/Bohr (for
residual forces) and 0.5 mHa/atom (for the tolerance on the
total energy). In these calculations, cutoff kinetic energies of
 35 Ha, 30 Ha and 35 Ha for the plane-wave basis set were used, and the Monkhorst-Pack sampling for the same tolerance criteria were determined to be $4\times4\times4$, $2\times4\times2$ and $2\times4\times4$, respectively, for YAG:Ce, LSO:Ce, and YAP:Ce, respectively.

The luminescence of Ce$^{3+}$ ions has been simulated in the supercell framework. 
The supercell sizes for YAG:Ce, LSO:Ce, and YAP:Ce are 80, 64 and 80 atoms, respectively, with Ce concentration of 8.33\%, 6.67\% and 6.67\%, respectively.  
Density-Functional Theory (DFT) calculations were performed with the PAW methodology and DFT+U (U=4.6~eV) to treat the 4f orbitals of Ce. 
In the excited state, we constrain the  eigenfunctions with dominant Ce$_{4f}$ character to be unoccupied, while occupying the next energy state higher in energy, which has been identified as being predominantly built from Ce$_{5d}$ orbitals for the three paradigmatic materials. 
We refereed to this technique as a “constrained DFT” approach in our earlier publication\cite{Jia-2019-2}
but the same term has been used in the quantum chemistry literature to denote a technique where the amount of charge in a given region of space is constrained.\cite{wu2005,wu2006,Kaduk2012,ramos2018}
The transition energy was calculated through the $\Delta$SCF method, that is, relying on total
energy differences of the different constrained configurations.

The stability of the $\Delta$SCF  method is known to depend on the investigated system and specific sought excitation state. 
Indeed, it can fail if the energy ordering of the different states varies with the SCF iteration.
In our previous works~\cite{jia-Ce,Jia-2019-2,Abinit2009,Gonze2016}, with the electronic state at the ground-state and excited-state geometries, this situation was never met. However, the situation is different with the OCP method, see the end of this section.

When the electronic one-electron eigenvalues come close to be degenerate, the $\Delta$SCF  method can not be used anymore, due to this problem. 
Our current implementation of the OCP method fails in the case of degenerate eigenstates. 
However, we do not think
that the OCP method per se is invalidated, as there are other ways to compute approximate excited state energies including near-degeneracy situations.\cite{ramos2018} The present work will be exemplified in the region far from crossing, where nevertheless (i) important insights about the possible barrier lowering is obtained, (ii) an estimation of the barrier will be inferred thanks to extrapolation, and (iii) the EDPH approach will be validated. 

We compute the OCP as a function of $\Lambda$, obtain the corresponding $\Delta E$ and the different quantities mentioned in Sec.~\ref{sec:theory}. More precisely, for each $\Lambda$, we start from some atomic geometry configuration, we compute the forces and stresses of the doped supercell for ground and excited state of the Ce$^{3+}$ ion, and combine
them in the OCP following Eq.~(\ref{eq:F}). Optimization on the basis of such forces and stresses has been implemented in the ABINIT software v8.10 (see the documentation: \texttt{imgmov}=6, the input variable \texttt{mixesimgf} controls the $\Lambda$ value). At the end, we obtain for this $\Lambda$, the relaxed geometry, as well as the corresponding $\Delta E$. 
There is no intrinsic limitation for the excited state characterization in the OCP methodology, thus we expect our approach to be useful to study other broad-band luminescent materials, such as the Eu$^{2+}$-doped phosphors, as well as the auto-ionization process for the thermal quenching behavior. 

However, the $\Delta$SCF approach is not always stable, due to changing energy ordering of the orbitals during the SCF cycle. 
For the three materials, we have observed such failure whenever the $\Delta$E value is below some threshold, depending on the material. 
Such threshold even varies according to the specific trajectory that is present in the OCP method, as will be illustrated in the next section.

\section{Results and Discussion}
\label{result}

\subsection{Comparing the 1D-CCM and the DEPH}

As a first step, we compute the total energy of the electronic ground and excited states at the relaxed ground-state ($Q_g$) and excited-state ($Q_e$) geometries, see Sec.~{\ref{sec:basic}}, then deduce the E$_{fd}$ energy
from the DEPH, according to Sec.~{\ref{sec:DEPH}}.
This is presented in Table~\ref{tab:Ce-table}. The smallest E$_{fd}$ energy is found for YAP:Ce (1.59 eV), while the largest one is found for LSO:Ce (4.40 eV). 
Even though these are relevant only for a semi-classical estimation of the non-radiative recombination, especially its temperature dependence, such large energy barrier would prevent any relevant non-radiative recombination at room or moderately high temperature. This was already mentioned in 
Ref. \onlinecite{jia-Ce}, with the associated conclusion that the alternative auto-ionization mecanism was likely to dominate over the 4f-5d crossing mechanism to explain the non-radiative recombination.

\begin{table}[h]
\centering
\renewcommand\arraystretch{1.5}
\caption{First-principles values for absorption and emission energies, as well as Franck-Condon shifts, for three Ce$^{3+}$-doped phosphors.
These values deliver an estimation of the 
energy barrier E$_{fd}$ for 4f-5d crossing thanks to the double energy parabola hypothesis, Eq.~({\ref{eq:Eb}}). $\Delta$C is the change of curvature, Eq.~({\ref{eq:deltaC}}). All values in eV.}
\label{tab:Ce-table}
\begin{tabular}{ccccccc}\hline \hline
 & E$_{abs}$ &E$_{em}$ & E$_{FC,g}$ & E$_{FC,e}$ & $\Delta$C & E$_{fd}$ \\ \hline
Y$_3$Al$_5$O$_{12}$:Ce & 2.78 & 2.36 & 0.22  & 0.20   & -0.02   & 2.78      \\
Lu$_2$SiO$_5$:Ce & 3.80 & 3.32 & 0.26  & 0.22   & -0.04   & 4.40      \\
YAlO$_3$:Ce & 4.14 & 3.56 & 0.38  & 0.20   & -0.18   & 1.59      \\ \hline \hline
\end{tabular}
\end{table}

With the DEPH, one obtains actually the full behaviour of the total energies (ground and excited states) as a function of the underlying one-dimensional parameter, which can be compared with the ones from the 1D-CCM. This is presented in Fig. 2. 
While the DEPH functions are analytic, and can thus be represented whatever the value of the one-dimensional parameter, we meet instability problems with the 1D-CCM, beyond some value of the mixing parameter $x$ defined in Eq.({\ref{eq:x}}). 
These instability problems are due to the $\Delta$SCF method, and have been mentioned in Sec. {\ref{sec:FP}}.
This happens around $x=4$ for YAG:Ce, $x=3.5$ for LSO:Ce, $x=5$ for YAP:Ce.
Other techniques to predict the excited state energy, e.g. the Bethe-Salpeter equation~\cite{Onida2002}
might be more stable.
However, excited calculations with alternative methods are usually considerably more expensive than $\Delta$DFT ones.

The 1D-CCM and the DEPH results match well in the YAG:Ce case, while the deviation of the 1D-CCM from the DEPH results is noticeable for the LSO:Ce case, and even larger for the YAP:Ce case. 
In the LSO:Ce, the 1D-CCM energies are larger than the DEPH ones, thus clearly showing an anharmonic behaviour, while the crossing point apparently happens at a lower value of $x$, but at nearly the same total energy. 
For the YAP:Ce case, the ground-state energy is similar in the 1D-CCM and DEPH cases, but there is considerable deviation of the 1D-CCM from the DEPH case for the excited-state energy: the crossing does not happen even at $x=5$ in the 1D-CCM case, while it is predicted to occur before $x=4$ in the DEPH case. 
Also, E$_{fd}$ from the 1D-CCM is much larger (above 5 eV) than from the DEPH for this material.

One might thus be tempted to deduce that the DEPH provides a lower bound to the E$_{fd}$ in these materials, that is better predicted by the 1D-CCM. 
We will see that, in the range of validity of the OCP, on the contrary, the DEPH has more predictive power than the 1D-CCM, and that it provides an upper bound of the energy barrier for our three materials with respect to OCP. 
Actually, the hypothesis of linear mixing of Q$_g$ and Q$_e$ happens to be a strong constraint to find the configuration that provides the lowest ground-state energy for a fixed absorption energy.

To end this subsection, we also provide in Fig.~{\ref{LSO-1D}} the electronic band structure of LSO:Ce in the excited state as a function of $x$ in the 1D-CCM. 
The $4f$ manifold energy (below 2 eV for $x=0$) gradually increases with increasing $x$, while the energy of the $5d$ band (around 4.8 eV for $x=0$) gradually decreases.

\subsection{Comparing the OCP with the 1D-CCM and the DEPH}

The 1D-CCM and OCP first-principles ground and excited state energies for YAG:Ce, LSO:Ce and YAP:Ce, as a function of the $\Delta$E variable, are shown in Fig.~\ref{figure_Delta}, and compared with the DEPH from Eq.~(\ref{eq:EDeltaE}). 
Note that the behaviour of $\Delta$E (horizontal axis) is opposite to the behaviour of the $x$ variable of Figs.~\ref{ccd} or~\ref{figure_1D}: the 4f-5d crossing is obtained on the right side of 
Figs.~\ref{ccd} or~\ref{figure_1D}, but on the left side of Fig.~\ref{figure_Delta} (when $\Delta$E=0). 
Moreover, in Fig.~\ref{figure_Delta} we show the range of energy differences 
down to the lowest energy where the $\Delta$SCF method fails to find the excited state in all the cases (1D-CCM or OCP). 
The $\Delta$SCF method is never stable at $\Delta$E=0.

One can easily distinguish the 1D-CCM and DEPH behaviors in  Fig.~\ref{figure_Delta}. 
While both reasonably match in the YAG:Ce case, the 1D-CCM points being slightly higher than the DEPH curves, the 1D-CCM results are clearly larger than the DEPH results for the LSO:Ce case, and even more so for the YAP:Ce case. 
This is in line with the analysis of the previous section.

Turning now to the OCP results,
the local minimum problem has to be accounted for in Fig.~\ref{figure_Delta}.  
Indeed, different local minima can be found using different geometries as starting point for a given value of $\Lambda$. 
We report the total energies of configurations reached by starting a local optimization algorithm from three different geometries, namely Q$_g$, Q$_e$ and another, test, geometry (see the caption).
The corresponding results are indicated by different symbols in Fig.~\ref{figure_Delta}.

Taking the OCP results as a whole (without distinguishing which starting point was used to generate them), one observe the following, for the three materials.
In the YAG:Ce case, the OCP results closely follow the DEPH results, thus being slightly lower than the 1D-CCM results. 
The latter is an expected behaviour, since the OCP searches for the lowest ground-state energy for a given energy difference. 
In the LSO:Ce case, the OCP results are also closer to the DEPH results than the 1D-CCM results. 
Still, in this case, the OCP results are lower than the DEPH, the deviation being on the same order of magnitude than the 1D-CCM one (but of opposite sign).
In the YAP:Ce case, we observe that the correspondence between the OCP results and the DEPH ones is excellent. 
In view of the strong deviation of the 1D-CCM from the DEPH results, such an excellent agreement is rather unexpected.
We can only observe that this agreement between the OCP and the DEPH implies that there is a path in configuration space for which the behaviors of the ground-state energy and the one of the excited-state energy are very parabolic, while, at variance, the very path must be rather non-linear, since otherwise it would be in excellent agreement with the 1D-CCM.

Of course, one cannot guarantee that the OCP global minimum was reached, whatever the material, whatever the $\Lambda$ value.
However, such local minima problem does not seem to affect the outcome of the present study:
for the three materials, the OCP results reasonably agree with the DEPH, in the following regions for $\Delta$E, decreasing from $E_{abs}$: 2.8~eV-1.4~eV (YAG:Ce), 3.8~eV-1.0~eV (LSO:Ce) and 4.1~eV-1.0~eV (YAP:Ce).

We now focus on the three points obtained from the OCP with initial Q$_g$ geometry for the LSO:Ce, in the $\Delta$E region of 1.0~eV-0.5~eV, Fig.~\ref{figure_Delta}. These points largely deviate from DEPH results. 
Table~\ref{tab:LSO-table} lists the typical OCP total energies and $\Delta$E in LSO:Ce. 
The problematic three points in LSO:Ce are from $\Lambda=1.5, 1.6, 1.7$, using Q$_g$ geometry as starting point.  
The band structures of LSO:Ce are shown in Figure~\ref{figure_bs}. 
With Q$_e$ geometry as starting point, the obtained OCPs indicate a much smaller geometry relaxation than that of Q$_e$ geometry. 
As a results, the band structures of LSO:Ce with $\Lambda=0-1.5$ from Q$_e$ geometry show a small change for the Ce states with respect to the band edge, while the results of $\Lambda=1.5$ from Q$_g$ geometry clearly depict a large difference. 
The Ce$_{5d}$ state does not appear inside the band gap in this case.
The deviation from the DEPH is thus due to the different nature of the electronic state. Another minimum of the OCP functional is found, but does not correspond to a $5d$ excited state.
Hence, such OCP result should not be included in the study of the 4f-5d crossing mechanism.

\begin{table}[h]
\centering
\renewcommand\arraystretch{1.5}
\caption{OCP energies and $\Delta$E in Lu$_2$SiO$_5$:Ce, for four
different $\Lambda$ values, using Q$_e$ or Q$_g$ geometry as starting point. The energies are quoted with respect to E$_{gs}$ at $\Lambda$=0, and expressed in eV. }
\label{tab:LSO-table}
\begin{tabular}{cccc}\hline \hline
 & E$_{gs}$ &E$_{ex}$ & $\Delta$E \\ \hline
$\Lambda$=0, Q$_e$ & 0.00 & 3.80 & 3.80   \\
$\Lambda$=0.5, Q$_e$ &  0.02 & 3.67 & 3.65  \\
$\Lambda$=1.0, Q$_e$ &  0.26 & 3.58 & 3.32  \\
$\Lambda$=1.5, Q$_e$ & 0.33 & 3.61 & 3.28 \\
$\Lambda$=1.5, Q$_g$ & 3.64 & 4.46 & 0.82   \\
\hline \hline
\end{tabular}
\end{table}

\begin{figure*}
\includegraphics[scale=0.45]{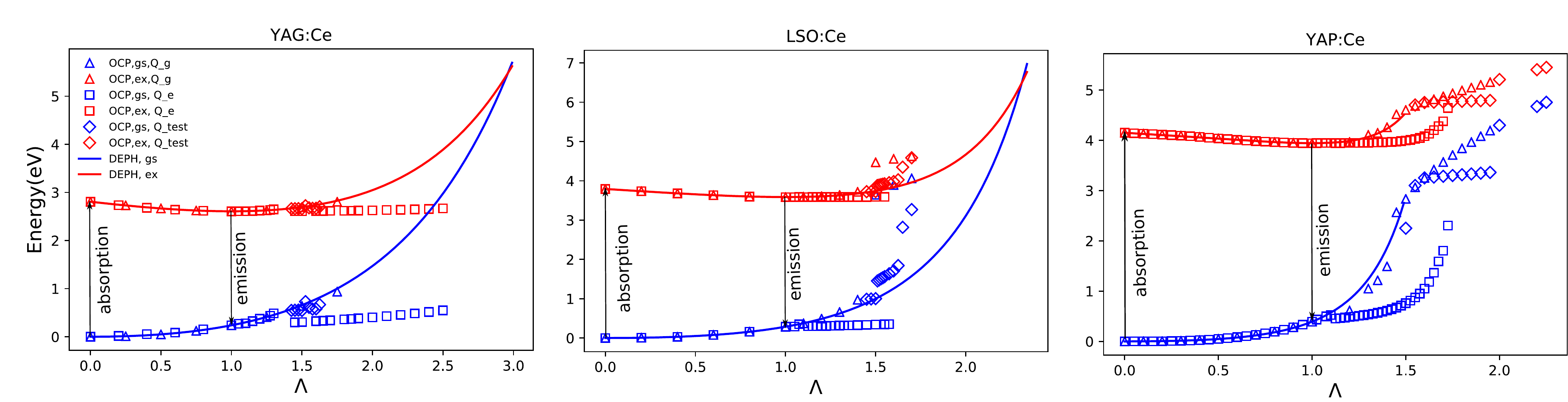}
\caption{The ground-state and excited-state total energies as a function of $\Lambda$, from the OCP method and from the DEPH, in three Ce$^{3+}$
-doped phosphors.(a) Y$_3$Al$_5$O$_{12}$:Ce (b) Lu$_2$SiO$_5$:Ce; (c) YAlO$_3$:Ce. The DEPH continuous lines correspond to Eqs.~(\ref{eq:EDEHPLambda}) and (\ref{eq:E*DEHPLambda}) with three parameters (per material) determined by first-principles data (i.e. they are not fitted to the OPC method points).
}\label{figure_fit}
\end{figure*}

The discussion of the OCP method  is now complemented by a comparison between the values obtained for the same Lagrange parameter $\Lambda$, for different starting points.
This is shown in Fig.~\ref{figure_fit}.
At variance with the nice agreement between OCP and DEPH in the Fig.~\ref{figure_Delta}, the representation of the ground and excited state between OCP and DEPH as a function of $\Lambda$ is more subtle to analyze. 
In the range of $\Lambda$ between 0 and 1, all the OCP energy points agree, and match very well the DEPH curve.
However, the situation changes when the $\Lambda$ value is above 1, the points depend strongly on the starting geometry used. 
For the three materials, the results from Q$_e$ and DEPH widely differ, while the results from Q$_g$ and Q$_{test}$ match with DEPH in a limited range of $\Lambda$. 
This might seem odd, but simply comes from the fact that the Lagrange parameter is a mathematical auxiliary quantity, that has no physical meaning. 
There is no obvious one-to-one correspondance between $\Lambda$ values from different configuration trajectories, even if they deliver the same $\Delta$E values.

\subsection{Ruling out the 4f-5d crossing as thermal quenching mechanism}

Finally, we discuss the thermal quenching mechanism of the three Ce$^{3+}$-doped phosphors.
In the present work, we have not focused on the accurate comparison between the 4f-5d crossover and auto-ionization models for thermal quenching,\cite{ponce-2016,Jia-2019-2} but on the description, validation, and testing of the OCP method, and on the size of the 4f-5d energy barrier. 
Indeed, the obtained results provide valuable information already at the level of the agreement between OCP and DEPH results. 
Such agreement shown in Fig.~\ref{figure_Delta}, validates the DEPH, and can then be extrapolated using the latter hypothesis. 
Thus, the energy barrier for the 4f-5d crossover from Eqs.~(\ref{eq:Eb}) or (\ref{eq:Eb2}) is indeed meaningful, and is much larger than the experimental thermal quenching barrier of 0.81~eV (YAG:Ce), 0.32~eV(LSO:Ce) and 1.20~eV(YAP:Ce),\cite{lyu1991,Peak2011} as listed in Table~\ref{tab:Ce-table}. 
This conclusion confirms our previous prediction of the irrelevance of the 4f-5d thermal quenching behaviour of Ce$^{3+}$-doped phosphors.\cite{Jia-2019-2} 
The probable thermal quenching mechanism may be the auto-ionization of an electron or the one of a hole. 
However, a solid assessment of such auto-ionization mechanism is left for a future study.

We note that free carrier (hole) non-radiative recombination rates in semiconductors have been computed on the basis of the 1D-CCM hypothesis for other materials, with good agreement with experiment.\cite{Walle2018} 
We expect that such calculation can also be conducted for auto-ionization process in white LED phosphors, even when this hypothesis is not true anymore, and the barrier might be significantly lower, as shown in the present work. 
However, the methodology for doing such calculation is to be established. 
The analysis of non-radiative rates should also include the discussion of symmetry related issues. These are present in matrix elements of the non-radiative $5d-4f$ transition, driven by electron-phonon coupling, but are not addressed in the present work.


\section{Conclusion}
\label{conclusion}

In this work, we propose a new methodology to explore the configuration space beyond the one-dimensional configuration model (1D-CCM). 
The latter focuses on linear configuration changes going from the equilibrium ground-state configuration to the equilibrium excited-state configuration. 
In the new methodology, a  Lagrange multiplier approach defines an optimal configuration path (OCP) in the configuration space, that includes both equilibrium configurations, but also extents to the ground-state excited-state crossing point, when it exists.
The ground-state and excited-state forces are mixed instead of their configurations like in the 1D-CCM.
The consequences of a double energy parabola hypothesis (DEPH) have also been worked out, valid for any configuration space trajectory that includes the equilibrium configurations. 
Three  paradigmatic phosphors, Y$_3$Al$_5$O$_{12}$:Ce, Lu$_2$SiO$_5$:Ce and YAlO$_3$:Ce were chosen to test the new method. In these phosphors, the luminescent mechanism is due to 4f-5d excitation.
One of the proposed thermal quenching mechanisms involves 4f-5d crossing. 
The resulting energy landscape is a function of the difference between ground-state and excited-state energies and was compared between OCP and DEPH. 
An agreement between both methods has been found, in the range of Lagrange parameters for which a distinct 5d excited state exists. 
This validates the DEPH for the  Ce$^{3+}$-doped phosphors.
%
By contrast, we observe a larger over-estimation energy barrier of 4f-5d crossing from the 1D-CCM approach. 
The OCP approach succeeds to find configurations with lower energies for the same energy difference than the 1D-CCM. 
The 4f-5d crossing are excluded as the thermal quenching mechanism of white LED phosphors in the three materials that we have investigated.

\begin{acknowledgments}
We acknowledge the help of J.-M. Beuken and M. Giantomassi for computational matters.
This work, done in the framework of ETSF (project number 551), has been supported by the Fonds de la Recherche Scientifique (FRS-FNRS Belgium) through a Charg\'e de recherches fellowship (Y. Jia) and the PdR Grant No. T.0103.19 - ALPS (X. Gonze). 
Computational resources have been provided by the supercomputing facilities of the Universit\'e catholique de Louvain (CISM/UCL) and the Consortium des Equipements de Calcul Intensif en F\'ed\'eration Wallonie Bruxelles (CECI) funded by the Fonds de la Recherche Scientifique (FRS-FNRS Belgium) under Grant No. 2.5020.11.
The present research benefited from computational resources made available on the Tier-1 supercomputer of the 
F\'ed\'eration Wallonie-Bruxelles, infrastructure funded by the Walloon Region under the grant agreement No. 1117545.
\end{acknowledgments}

\end{document}